\documentstyle[11pt,aasms4]{article}
\begin{document}
\title{Oscillator Strengths for $B - X$, $C - X$, and $E - X$
Transitions in Carbon Monoxide}

\author{S. R. Federman\footnotemark[1], M. Fritts\footnotemark[1], 
S.  Cheng\footnotemark[1], K.M. Menningen\footnotemark[2], 
David C. Knauth\footnotemark[1], and K. Fulk\footnotemark[2]}

\footnotetext[1]{Department of Physics and Astronomy, University of Toledo, 
Toledo, OH  43606.}
\footnotetext[2]{Department of Physics, University of Wisconsin-Whitewater, 
Whitewater, WI 53190.}

\begin{abstract}

Band oscillator strengths for electronic transitions in CO 
were obtained at the Synchrotron Radiation Center 
of the University of Wisconsin-Madison.  Our focus was on 
transitions that are observed in interstellar spectra with the 
{\it Far Ultraviolet Spectroscopic Explorer}; these transitions are also 
important in studies of selective isotope photodissociation where 
fractionation among isotopomers can occur.  Absorption from the ground state 
($X$ $^1\Sigma^+$ $v^{\prime\prime}$ $=$ 0) to 
$A$ $^1\Pi$ ($v^{\prime}$~$=$~5), $B$ $^1\Sigma^+$ ($v^{\prime}$~$=$~0,~1), 
$C$ $^1\Sigma^+$ ($v^{\prime}$~$=$~0,~1), and 
$E$ $^1\Pi$ ($v^{\prime}$~$=$~0) was measured.  
Fits to the $A - X$ (5,~0) band, whose oscillator strength is 
well known, yielded the necessary column density and excitation temperature.  
These parameters were used in a least-squares fit of the 
observed profiles for the transitions of 
interest to extract their band oscillator strengths.  Our oscillator strengths
are in excellent agreement with results from recent experiments using a 
variety of techniques.  This agreement provides the basis for 
a self-consistent set of $f$-values at far ultraviolet 
wavelengths for studies of interstellar (and stellar) CO.
\end{abstract}


\newpage
\section{Introduction}

Carbon monoxide, the second most abundant interstellar molecule, is seen in 
cloud envelopes through absorption at ultraviolet wavelengths.  At wavelengths 
above 1200 \AA, spectra acquired with the {\it Hubble Space Telescope} ($HST$) 
revealed CO absorption from the dipole allowed $A - X$ series of bands 
(Sheffer et al. 1992; Lambert et al. 1994; Lyu, 
Smith, \& Bruhweiler 1994) and from several 
weak triplet-singlet bands (Federman et al. 1994).  At shorter wavelengths, 
the (0, 0) bands of the allowed $B - X$, $C - X$, and $E - X$ transitions 
were seen in spectra taken with the $Copernicus$ satellite (e.g., Morton 
1975; Snow 1975; Federman et al. 1980).  The recently launched 
{\it Far Ultraviolet Spectroscopic Explorer} provides the means to study 
these short-wavelength transitions again.

Oscillator strengths are required to convert the observed amount of 
absorption into abundances or column densities.  Recent experimental work, 
using a variety of techniques, has produced a set of self-consistent 
$f$-values for the bands at 1200 \AA\ and beyond.  In particular, the results 
of Chan et al. (1993), Smith et al. (1994), Federman et al. (1997), Jolly 
et al. (1997), Zhong et al. (1997), Stark et al. (1998), and Eidelsberg et al. 
(1999) for $A - X$ ($v^{\prime}$ $=$ 5-16) agree within experimental 
uncertainties, typically 10\%.  As for the triplet-singlet bands, the analysis 
of Rostas et al. (2000) comfirms the $f$-values inferred from $HST$ 
observations by Federman et al. (1994).  The accuracy of these $f$-values now 
is also 10\%.

The situation for transitions involving the Rydberg states, $B$, $C$, and 
$E$, where factor-of-2 differences persist, is less satisfactory.  The 
results of earlier experiments based on CO absorption (Eidelsberg 
et al. 1991; Stark et al. 1992) tended to be smaller than those $f$-values 
obtained in electron-energy-loss and electron-impact-excitation 
experiments (Chan et al. 1993; Kanik et al. 1995; Ciocca et al. 1997; 
Zhong et al. 1997).  Theoretical determinations (Kirby \& Cooper 1989; 
Chantranupong et al. 1992; Rocha, Borges, \& Bielschowsky 1998) usually 
favored the smaller values, but spanned a factor-of-2 range.  In this paper, 
we present new results based on absorption of synchrotron radiation.  Our 
$f$-values are consistent with the results of experiments using electrons and 
with the recent work of Stark et al. (1999) on $B - X$ (0, 0) and (1, 0), 
which involved absorption of laser light.  The combined effect is that 
interstellar studies now have a reliable set of CO $f$-values covering the 
range 1050 to 1550 \AA.

These results impact other facets of CO spectroscopy.  First, the destruction 
of CO in interstellar space and circumstellar envelopes involves line 
absorption into predissociating levels (e.g., Bally \& Langer 1982; 
van Dishoeck \& Black 1988; Warin, Benayoun, \& Viala 1996).  
All states lying above the $E$ ($v$ $=$ 1) state predissociate.  
Absorption into the $E$ ($v$~$=$~0,~1) and $C$ ($v$~$=$~1) states 
leads to photodissociation a significant fraction of the time.  The 
photodissociation rate depends on the optical depth of the absorption line, 
which in turn depends on the line oscillator strength.  Use of larger 
$f$-values, as suggested by our work and others, will alter model 
predictions.  Second, once oscillator strengths are known, measurements of 
the branching fraction for transitions to the ground state reveal upper state 
lifetimes (e.g., Cacciani et al. 1998).  We obtained lifetimes 
from our $f$-values and available branching fractions and compared these with 
information in the literature.

\section{Experiment}

Our absorption experiment was conducted at the Synchrotron Radiation Center 
of the University of Wisconsin-Madison.  The setup is shown in Figure 1.  
Ultraviolet (UV) light was collimated through the 
entrance slit of the 4 m Normal Incidence Monochromator 
and passed through a LiF window before 
entering a CO gas cell that is 14.6 cm long.  At the end of the gas cell, 
the UV light was converted into visible light by a sodium salicylate coating 
on a flange window.  The light was 
focused by a lens onto the cathode of the photomultiplier tube (PMT) 
(Hamamatsu R3788).  The photocurrent from the PMT was converted into 
voltage and sampled by an analog-to-digital converter.  The beam current 
from the synchrotron and the total pressure inside the gas cell were 
continuously monitored.  A Pirani ion gauge was used for pressures above 
1.3 mTorr, and a cold cathode ionization gauge was used for lower pressures.  
Slit widths of 15 and 20 $\mu$m were 
used to maximize the spectral resolution, yielding $\delta\lambda$ of $\sim$ 
0.11 \AA.  The total gas pressure ranged from 0.3 
to 18.3 mTorr, corresponding to CO column densities of 
$8 \times 10^{13}$ to $5 \times 10^{15}$ cm$^{-2}$, 
values that are observed in interstellar clouds.  In the analysis that 
follows, we used column densities derived from synthesized spectra of 
measurements on the $A - X$ (5,~0) band instead of pressures to minimize 
systematic effects arising from impurities contributing to the pressure and 
from measurements with two pressure gauges.  Data were acquired on $B - X$ 
(0,~0), $B - X$ (1,~0), $C - X$ (0,~0), $C - X$ (1,~0), and $E - X$ (0, 0) 
bands.  Wavelength scans at every band for each pressure in the cell were 
followed by a scan at $A - X$ (5, 0) at the same pressure.  The data for the 
$A - X$ (5, 0) band were used to place our results for the Rydberg bands 
on an absolute scale.  In most instances several scans were obtained 
for a specific pressure.  The step size was set either to 0.01 or 0.02 \AA\ 
for a scan; no systematic effects on the derived $f$-values were noticed for 
scans with different step sizes.

The transmission curve of the system is given in Figure 2, with the 
CO bands superposed.  In addition to the bands analyzed here, several 
other features are present in Fig. 2.  First, there is a contaminant, most 
likely water, with features between 1200 and 1250 \AA\ and near 1125 \AA.  
While the latter occurs where the $B - X$ (1, 0) band lies, it is sufficiently 
broad that no problems arose in extracting the band of interest.  At 
the higher pressures, the $B - X$ (2, 0) and $E - X$ (1, 0) bands were seen.  
The $B - X$ (2, 0) band was always too weak to derive a reliable $f$-value 
from the data.  Since an accurate measurement of the $A - X$ (5, 0) band 
was necessary, the pressure could not be set arbitrarily high to get good 
data on $B - X$ (2, 0).  
While the signal to noise was better for the $E - X$ (1, 0) 
band, no reliable determinations were possible because this band lies too 
close to the LiF cutoff.

\section{Analysis and Results}

The data for each band was processed in much the same way as in 
Federman et al. (1997).  The dark current from the PMT, which was measured 
before each scan by closing a mechanical shutter to the beamline, was 
subtracted from the signal and the 
beam current drift was removed by normalizing to the beam current curve 
for the scan.  The data were corrected for the 
transmission curve of the setup, which was obtained 
with no CO in the cell.  Spectra taken at a given pressure were combined.  
The flux from the synchrotron was fit to a low order polynomial, yielding 
normalized spectra for each pressure.  In some instances, a Gaussian fit 
to the contaminant feature near the $B - X$ (1, 0) band was needed to aid 
in its removal.  The latter 3 steps utilized the NOAO/IRAF package.

The next steps involved the spectra of the $A - X$ (5, 0) band.  
Each normalized spectrum was synthesized with a least-squares fitting 
routine (cf. Lambert et al. 1994) used by us in our earlier CO experiment 
(Federman et al. 1997).  In essence, the routine minimizes the difference 
between measured and synthesized amounts of absorption.  
The synthetic spectrum was based on line wavelengths from 
Tilford \& Simmons (1972) and the well-determined $f$-value of 
$1.45 \times 10^{-2}$ (e.g., Chan et al. 1993).  Voigt line profiles 
were adopted for all syntheses.  Excitation temperature, 
wavelength offset, column density, and instrumental width 
(well approximated by a Gaussian) were varied.  
Once the instrumental width for each of the two slit-widths ($\sim$ 0.11 \AA) 
was determined, the spectra were resynthesized, giving the column density 
for a specific pressure.  This column density was used in the fit of the 
Rydberg bands.  As in our earlier work, the excitation temperature 
was usually less than room temperature.  This difference arises because 
the high-lying rotational levels decay radiatively at the pressures in the 
gas cell.  We investigated the effects of changing the excitation temperature 
on the derived $f$-values for the Rydberg bands; the $f$-values changed by 
1\% or less.  For excitation temperatures in the range 250 to 300 K, the 
absorption was spread over many rotational levels, thereby reducing the 
effect of optical depth on individual lines.

The determination of the $f$-values for Rydberg bands at various pressures 
was accomplished through profile synthesis, varying only the 
wavelength offset and $f$-value.  Here the line lists of Eidelsberg et al. 
(1991) were used because they were the most complete.  The wavelengths for 
the $B - X$, $C - X$, $E - X$ bands studied by us were not affected by 
the recalibration described in Eidelsberg et al. (1992).  The initial 
$f$-values in the fitting process were those listed by Morton \& Noreau 
(1994).  Typical fitting results for each band 
are presented in Figure 3.  For the strong $C - X$ (0, 0) band, the data for 
a pressure of 0.6 mTorr are shown; all other fits are to data taken at 
3 mTorr.

The uncertainty in the inferred oscillator strengths includes a statistical 
component and a systematic one.  The 
statistical error in the $f$-value for a band at 
a specific pressure was obtained by summing in quadrature 
the error in the equivalent width ($W_{\lambda}$) of the reference band, 
the error in $W_{\lambda}$ of the band in question, and the error in the 
reference band $f$-value, taken to be 5\% (Chan et al. 1993).  
A conservative uncertainty in $W_{\lambda}$ 
was determined by multiplying the root-mean-square variation in 
the continuum flux from the synchrotron by the width of the profile 
at half-maximum depth.  Systematic effects included 
variations in pressure for a given set of measurements ($\le$ 5\%), 
uncertainties in instrumental width ($\le$ 5\%) and in 
excitation temperature ($\le$ 1\%), and the problems fitting the Q branch for 
$E - X$ (0, 0) ($\le$~5\%), as noted in the next section.  The combined 
effect introduces a maximum uncertainty of 8.7\% in our derived oscillator 
strength for the $E - X$ (0, 0) band and 7.1\% in the others.

The experimental results for each absorption 
band are summarized in Table 1. The error bars quoted there are due to 
statistics only.  Results are only shown for pressures where the signal to 
noise in the continuum allowed a meaningful determination and where problems 
associated with optically thick lines were kept to a minimum.  The excellent 
agreement in derived $f$-value from one pressure to another gives us 
confidence in the results.  At the largest pressures for a given band, the 
optical depth at line center for the strongest line in the $B - X$ and 
$C - X$ bands is modest ($\le$ 6).  While the optical depth for the Q(7) 
line of the $E - X$ (0, 0) transition reaches nearly 40, the results are 
consistent for all pressures with optical depths from 0.9 to $\sim$ 40.  
A comparison of our results with other 
experimental and theoretical results is presented in Table 2.  Here the 
uncertainty includes systematic errors as well.

\section{Discussion}
\subsection{Fits to Rydberg Bands}

Several comments about the fits to the Rydberg bands (Fig. 3) deserve 
mention.  In all cases, fits to the P and R branches show less absorption 
for the high-lying rotational levels than is observed.  Changes in 
excitation temperature do not remedy the situation.  A more likely cause 
lies in the fact that our fits are restricted to lines for levels up to 
$J$ $=$ 25.  While Tilford \& Simmons (1972) included more levels in their 
compilation, their list for the Q branch of the $E - X$ 
(0, 0) band was incomplete.  
We felt that self-consistency in use of line list was more important than the 
slight deviations in profile synthesis for lines involving high-lying 
rotational levels in the P and R branches.  Moreover, the fraction of CO 
molecules in J $\ge$ 26 at an excitation temperature of $\sim$ 280 K is 
less than 0.2\%; absorption from these levels contributes less than about 
1\% to the total $W_{\lambda}$ for a band.  Such a small effect was not 
considered in the error budget.

Our fit to the Q branch of $E - X$ (0, 0) is less than satisfactory 
because the fit suggests more absorption than is seen in the laboratory data.  
Eidelsberg et al.'s (1991) line list 
for the Q branch is based mainly on theoretical estimates because the band 
was not resolved.  Recent laser experiments (Cacciani, Hogervorst, \& Ubachs 
1995) indicate that the rotational constant for the $e$-parity 
$\Lambda$-doublet component (Q branch) is about 0.5\% larger than the one for 
the $f$ component (P and R branches).  This translates into a broadening of the 
Q branch, which would allow syntheses to track the measured profiles better.  
Since Cacciani et al. did not give a line list, we preferred to use the 
available information in Eidelsberg et al. (1991).  The measured wavelengths 
for the P and R lines in Eidelsberg et al. are consistent with other 
determinations; typical differences of 3 m\AA\ with Tilford \& Simmons (1972), 
for instance, are seen.  
(Since our synthesis is based on all three branches, the resulting $f$-value 
for the band is not likely to be affected greatly.  A systematic uncertainty 
of at most 5\% applies to the effect noted here.)

\subsection{Oscillator Strengths}

Our results are in excellent agreement with 
the recent work of Ciocca et al. (1997), Zhong et al. (1997), 
and Stark et al. (1999).  Good agreement with Chan et al. (1993) and Kanik et 
al. (1995) is also seen.  For the (0, 0) bands, earlier absorption work 
by Eidelsberg et al. (1991) and Stark et al. (1992) 
gave significantly smaller $f$-values, 
but results of Eidelsberg et al. show consistency with the recent work for 
the relatively weak (1, 0) bands.  A systematic trend is seen in the results 
presented by Eidelsberg et al.: The discrepancy between their $f$-values and 
the more recent determinations increases with increasing band strength.  This 
trend suggests optical depth effects were not adequately treated in their 
analysis of CO absorption.  The results of Stark et al. (1992) may be affected 
by this problem as well.  Agreement with theoretical results is seen 
at the factor-of-2 level.  Rocha et al. (1998) showed that the differences 
between their theoretical oscillator strengths, those of Chantranupong et al. 
(1992), and the experimental ones of Zhong et al. (1997) mainly 
arise from the differences in Generalized Oscillator Strength (GOS) for small 
$K^2$, the square of the transferred momentum.  The fact that all experimental 
techniques now give oscillator strengths similar to the ones obtained by 
Zhong et al. indicates more theoretical work on the GOS at small $K^2$ is 
warranted.

The results presented here and from other recent experiments impact the 
models of CO photodissociation.  According to the models of van Dishoeck \& 
Black (1988), the $E - X$ (0, 0) band plays an important role.  Their 
modeling efforts are based on the oscillator strengths of Letzelter et al. 
(1987), later updated by Eidelsberg et al. (1991).  The recent experiments, 
including ours, indicate the $f$-values of Eidelsberg et al. for the strongest 
bands are too small, by nearly a factor of 2 for the $E - X$ (0, 0) band.  
The discrepancy appears for $f$-values greater than about $6 \times 10^{-3}$.  
All other bands found to be important in the photodissociation process by 
van Dishoeck \& Black have $f$-values greater than $7.5 \times 10^{-3}$, 
according to Letzelter et al. (1987).  We suspect that all these $f$-values 
have to be revised upward.  Such a revision would greatly influence model 
predictions.  In particular, lines become optically thick closer to a cloud's 
edge when $f$-values are larger.  The effect would be enhanced self-shielding 
of the most abundant isotopic variant, $^{12}$C$^{16}$O, leading to enhanced 
fractionation relative to other isotopomers.  Such an effect could explain 
the large column ratios, $N$($^{12}$C$^{16}$O)/$N$($^{13}$C$^{16}$O), seen 
toward $\zeta$ Oph (Lambert et al. 1994).  For the $E - X$ (0, 0) band, which 
has a substantial radiative decay channel, the photodissociation yield is 
somewhat lower as well (see Cacciani et al. 1998).

\subsection{Radiative Lifetimes}

With a self-consistent set of oscillator strengths for the Rydberg bands 
now available, radiative lifetimes for the upper state can be inferred 
from branching fractions for transitions to the ground electronic state.  
The relationship between band oscillator strength, branching fraction, 
and lifetime is given in Morton \& Noreau (1994):

\begin{equation}
(2 - \delta_{0\Lambda^{\prime\prime}})\ f_{v^{\prime}v^{\prime\prime}} = 
1.499 \times 10^4\ (2 - \delta_{0\Lambda^{\prime}})\ 
\lambda^2_{v^{\prime}v^{\prime\prime}}\ A_{v^{\prime}v^{\prime\prime}}, 
\end{equation}

\noindent where $\lambda$ is the wavelength of the transition in meters 
and $A_{v^{\prime}v^{\prime\prime}}$ is the transition probability in 
s$^{-1}$.  The branching fraction $\gamma$ and radiative lifetime 
$\tau_{v^{\prime}}$ are related by 

\begin{equation}
\gamma = A_{v^{\prime}v^{\prime\prime}}/\sum_{v^{\prime\prime}} 
A_{v^{\prime}v^{\prime\prime}} = A_{v^{\prime}v^{\prime\prime}}\ 
\tau_{v^{\prime}}
\end{equation}

\noindent with the lifetime in seconds.  For the $X$, $B$, 
and $C$ states, $\delta_{0\Lambda}$ is 1.  
Combining these facts leads to the expression 

\begin{equation}
f_{v^{\prime}v^{\prime\prime}} = 1.499 \times 10^4\ 
\lambda^2_{v^{\prime}v^{\prime\prime}}\ \gamma/\tau_{v^{\prime}}
\end{equation}

\noindent from which we can derive $\tau_{v^{\prime}}$, knowing $\gamma$.

We focus on the $B - X$ (0, 0) and $C - X$ (0, 0) transitions, which do not 
lead to absorption into predissociating levels.  Essentially all decays from 
the upper states $B$ and $C$ ($v^{\prime}$ $=$ 0) return the molecule either 
to the $X$ or $A$ states (Letzelter et al. 1987; Kirby \& Cooper 1989), with 
nearly all decays to ground involving $v^{\prime\prime}$ $=$ 0.  
Experimental $B - X$ branching fractions span the range 0.56 to 0.68 
(Dotchin, Chubb, \& Pegg 1973; Letzelter et al. 1987; Drabbels, Meerts, \& ter 
Meulen 1993), indicating respective lifetimes of 17 to 20 ns when using our 
$f$-value.  The early measurements, summarized by Krishnakumar \& Srivastava 
(1986), and those of Drabbels et al. (1993) yielded lifetimes that are too 
long by 1 to several standard deviations.  The measurements of Hart \& 
Bourne (1989) are the most consistent with available branching fractions and 
our $f$-values.  The corresponding $C - X$ branching fractions range from 
0.94 to 0.99 (Dotchin et al. 1973; Letzelter et al. 1987; Drabbels et al. 
1993), leading to a lifetime of $\sim$ 1.4 ns for the $C$ ($v^{\prime}$~$=$~0) 
state.  Again, many of the measurements indicate lifetimes that are 1- to 
2-$\sigma$ too long, but the results of Hesser (1968) and Chornay, King, \& 
Buckman (1984) are consistent with our inferred lifetimes.  The lifetime of 
Drabbels et al. is much too short.  The fact that many of the lifetime 
determinations are too long suggests radiative trapping may have adversely 
affected the results.  As for the $E - X$ (0, 0) transitions, Cacciani et al. 
(1998) provide a similar analysis; their results using the $f$-value of 
Chan et al. (1993) appear to be the most appropriate.

\section{Final Remarks}

New results for oscillator strengths of several Rydberg bands in CO were 
obtained from an absorption experiment using synchrotron radiation.  The 
results are in excellent agreement with other recent experimental 
determinations based on laser absorption, electron energy loss, and electron 
impact excitation.  These independent measures reveal that 
a self-consistent set of band $f$-values for the $B - X$ 
(0,~0) and (1,~0), $C - X$ (0,~0) and (1,~0), and $E - X$ (0,~0) 
transitions is now available for studies of interstellar (and stellar) CO.  
Such studies include observations with the {\it Far Ultraviolet Spectroscopic 
Explorer} and models of CO photodissociation.  
The $f$-values for the strongest transitions are larger than once thought. 
Theoretical models of interstellar chemistry using 
the new $f$-values will show more fractionation of 
$^{12}$C$^{16}$O relative to less abundant forms.  
Finally, combining available branching fractions for 
transitions between the upper and ground states 
with our $f$-values provided the means to obtain radiative lifetimes 
for the upper states.  For the $B$ and $C$ states with $v^{\prime}$ $=$ 0, 
shorter lifetimes are most consistent with the set of larger $f$-values.

\newpage

\acknowledgments  We thank Yaron Sheffer for the code used in profile 
synthesis.  The research was supported by NASA grants NAG5-4957, NAG5-6729, 
and NAG5-7754 to the University of Toledo and by NSF grant DMR-95-31009 to 
the Synchrotron Radiation Center.

\newpage
\begin{center}
{\large Table 1} \\
{\large Experimental Results} \\
\begin{tabular}{cccccc}  \hline\hline
Pressure &  &  & $f$-values ($\times$ 10$^{3}$) &  &  \\ \cline{2-6}
(mTorr) & $B - X$ (0, 0) & $B - X$ (1, 0) & $C - X$ (0, 0) & $C - X$ (1, 0) & 
$E - X$ (0, 0) \\ \hline
0.3 & $\ldots$ & $\ldots$ & $123 \pm 10$ & $\ldots$ & $66 \pm 6$ \\ 
0.6 & $\ldots$ & $\ldots$ & $122 \pm 11$ & $\ldots$ & $60 \pm 6$ \\ 
1.0 & $\ldots$ & $\ldots$ & $\ldots$ & $\ldots$ & $61 \pm 4$ \\ 
1.29 & $\ldots$ & $\ldots$ & $\ldots$ & $\ldots$ & $74 \pm 4$ \\ 
1.65 & $7.1 \pm 0.5$ & $0.76 \pm 0.35$ & $\ldots$ & $3.1 \pm 0.6$ & 
$69 \pm 4$ \\ 
3.01 & $6.7 \pm 0.5$ & $0.79 \pm 0.18$ & $\ldots$ & $2.9 \pm 0.6$ & 
$63 \pm 4$ \\ 
6.08 & $\ldots$ & $1.09 \pm 0.42$ & $\ldots$ & $\ldots$ & $\ldots$ \\ 
6.09 & $5.7 \pm 0.6$ & $0.70 \pm 0.41$ & $\ldots$ & $2.6 \pm 1.0$ & 
$72 \pm 5$ \\ 
6.78 & $6.7 \pm 0.5$ & $\ldots$ & $\ldots$ & $\ldots$ & $69 \pm 4$ \\ 
9.03 & $\ldots$ & $0.79 \pm 0.37$ & $\ldots$ & $\ldots$ & $\ldots$ \\ 
9.75 & $6.9 \pm 0.4$ & $0.68 \pm 0.15$ & $\ldots$ & $3.1 \pm 0.4$ & 
$70 \pm 4$ \\ 
12.1 & $\ldots$ & $0.92 \pm 0.22$ & $\ldots$ & $\ldots$ & $\ldots$ \\ 
12.5 & $6.4 \pm 0.6$ & $0.92 \pm 0.27$ & $\ldots$ & $3.0 \pm 0.6$ & 
$68 \pm 5$ \\ 
12.9 & $6.4 \pm 0.5$ & $0.96 \pm 0.22$ & $\ldots$ & $2.7 \pm 0.5$ & 
$75 \pm 5$ \\ 
18.1 & $6.7 \pm 0.5$ & $0.70 \pm 0.14$ & $\ldots$ & $\ldots$ & $\ldots$ \\ 
18.3 & $6.9 \pm 0.5$ & $0.89 \pm 0.18$ & $\ldots$ & $\ldots$ & $\ldots$ \\ 
&  &  &  &  &  \\ 
{\bf average} & ${\bf 6.7 \pm 0.2}$ & ${\bf 0.80 \pm 0.06}$ & 
${\bf 123 \pm 7}$ & ${\bf 3.0 \pm 0.2}$ & ${\bf 68 \pm 1}$ \\ \hline
\end{tabular}
\end{center}

\newpage
\oddsidemargin=-0.5in
\evensidemargin=-0.5in

\begin{center}
{\large Table 2} \\
{\large Comparison of Results} \\
\begin{tabular}{lccccc} \hline\hline
Reference &  &  & $f$-values ($\times$ 10$^{3}$) &  &  \\ \cline{2-6}
 & $B - X$ (0, 0) & $B - X$ (1, 0) & $C - X$ (0, 0) & $C - X$ (1, 0) & 
$E - X$ (0, 0) \\ \hline
Experiment: &  &  &  &  &  \\
{\bf Present results }$^{a}$ & ${\bf 6.7 \pm 0.7}$ & ${\bf 0.80 \pm 0.12}$ 
& ${\bf 123 \pm 16}$ & ${\bf 3.0 \pm 0.4}$ & ${\bf 68 \pm 7}$ \\ 
Lassettre \& Skerbele 1971\ $^{b}$ & $15 \pm 3$ & $2.0 \pm 0.4$ & 
$163 \pm 15$ & $7.0 \pm 0.6$ & $94 \pm 9$ \\ 
Eidelsberg et al. 1991\ $^{a}$ & $4.52 \pm 0.45$ & $0.72 \pm 0.07$ & 
$61.9 \pm 6.2$ & $2.77 \pm 0.28$ & $36.5 \pm 3.7$ \\ 
Stark et al. 1992\ $^{a}$ & $\ldots$ & $\ldots$ & $\ldots$ & $\ldots$ & 
$49.0 \pm 5.0$ \\ 
Chan et al. 1993\ $^{b}$ & 8.03 & 1.32 & 117.7 & 3.56 & 70.6 \\ 
Kanik et al. 1995\ $^{c}$ & $12 \pm 3$ & $\ldots$ & $154 \pm 41$ & 
$\ldots$ & $\ldots$ \\ 
Ciocca et al. 1997\ $^{c}$ & $\ldots$ & $\ldots$ & $\ldots$ & $\ldots$ & 
$70.8 \pm 18.4$ \\ 
Zhong et al. 1997\ $^{b}$ & $5.98 \pm 0.93$ & $\ldots$ & $114 \pm 14$ & 
$3.22 \pm 0.94$ & $64.2 \pm 8.1$ \\ 
Stark et al. 1999\ $^{d}$ & $6.5 \pm 0.6$ & $1.1 \pm 0.1$ & $\ldots$ & 
$\ldots$ & $\ldots$ \\ 
Theory: &  &  &  &  &  \\ 
Kirby \& Cooper 1989 & 2.1 & 0.3 & 118.1 & 1.8 & 49 \\ 
Chantranupong et al. 1992 & 5.08 & 0.52 & 64.7 & 4.9 & 27.4 \\
Rocha et al. 1998 & 4.8 & 0.43 & 89 & 2.9 & 49 \\ \hline
\end{tabular}
\end{center}

\noindent $^{a}$\ Synchrotron absorption.\newline
$^{b}$\ Electron energy loss.\newline
$^{c}$\ Electron impact excitation.\newline
$^{d}$\ Laser absorption.

\newpage
\oddsidemargin=0.0in
\evensidemargin=0.0in

\newpage

\begin{center}
{\vbox to6.75in{\rule{0pt}{6.75in}}
\includegraphics{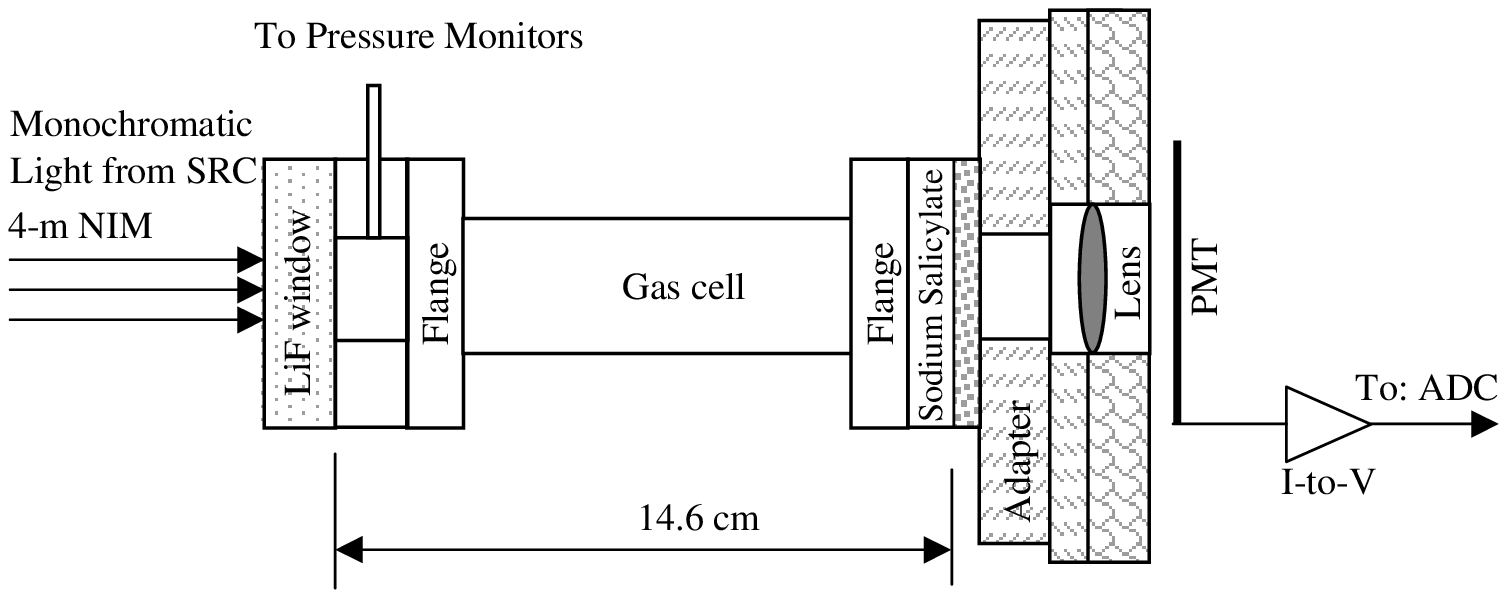}
\begin{flushleft}
Fig. 1 $-$ A schematic of the experimental setup at the 4 m Normal Incidence 
Monochromator beamline used for the CO measurements.
\end{flushleft}}
\end{center}

\newpage
\begin{center}
{\vbox to6.75in{\rule{0pt}{6.75in}}
\includegraphics{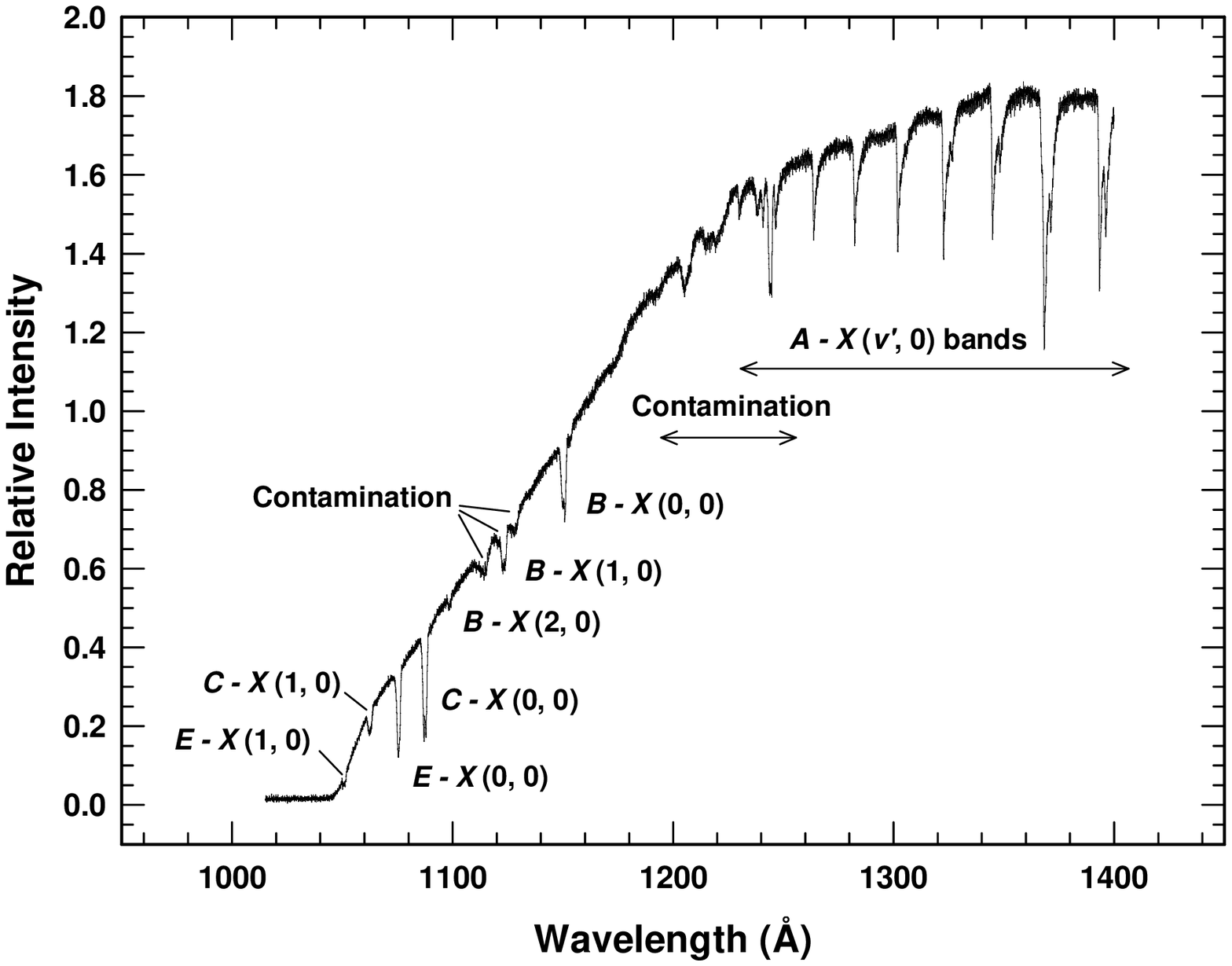}
\begin{flushleft}
Fig. 2 $-$ A scan showing the transmission curve of the LiF windows 
with a superposed spectrum of CO.  The CO bands are indicated, as are features 
from a contaminant, most likely H$_2$O.
\end{flushleft}}
\end{center}

\newpage

\begin{center}
{\vbox to5.80in{\rule{0pt}{5.80in}}
\includegraphics{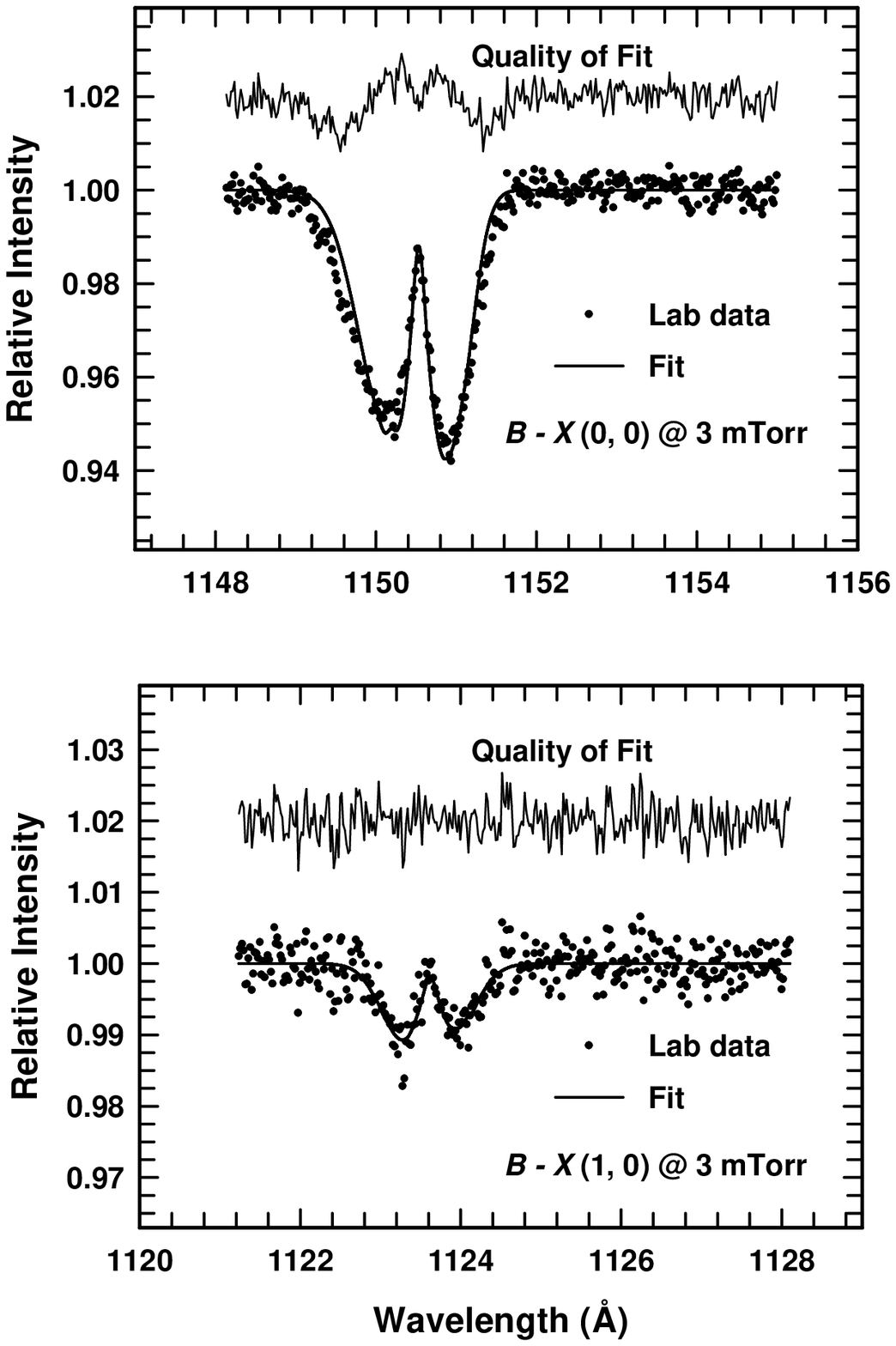}
\begin{flushleft}
Fig. 3 $-$ Fits to the CO bands. a) $B - X$ (0, 0) and (1, 0) bands. 
b) $C - X$ (0, 0) and \\
(1, 0) bands. c) $E - X$ (0, 0) and $A - X$ (5, 0) 
bands.  The strong $C - X$ (0, 0) band was taken at a pressure of 0.6 mTorr; 
all other data are from a pressure of 3 mTorr.  The data are represented by 
filled circles, the syntheses by thick curves, and the quality of the fit by 
thin lines.  The quality of the fit indicates the difference between fit and 
data, offset to 1.02.  The spectrum for the $C - X$ (1, 0) band has lower 
signal to noise, the result of lying near the LiF cutoff; here the quality of 
the fit is offset to 1.03.
\end{flushleft}}
\end{center}

\newpage

\begin{center}
\vbox to5.80in{\rule{0pt}{5.80in}}
\includegraphics{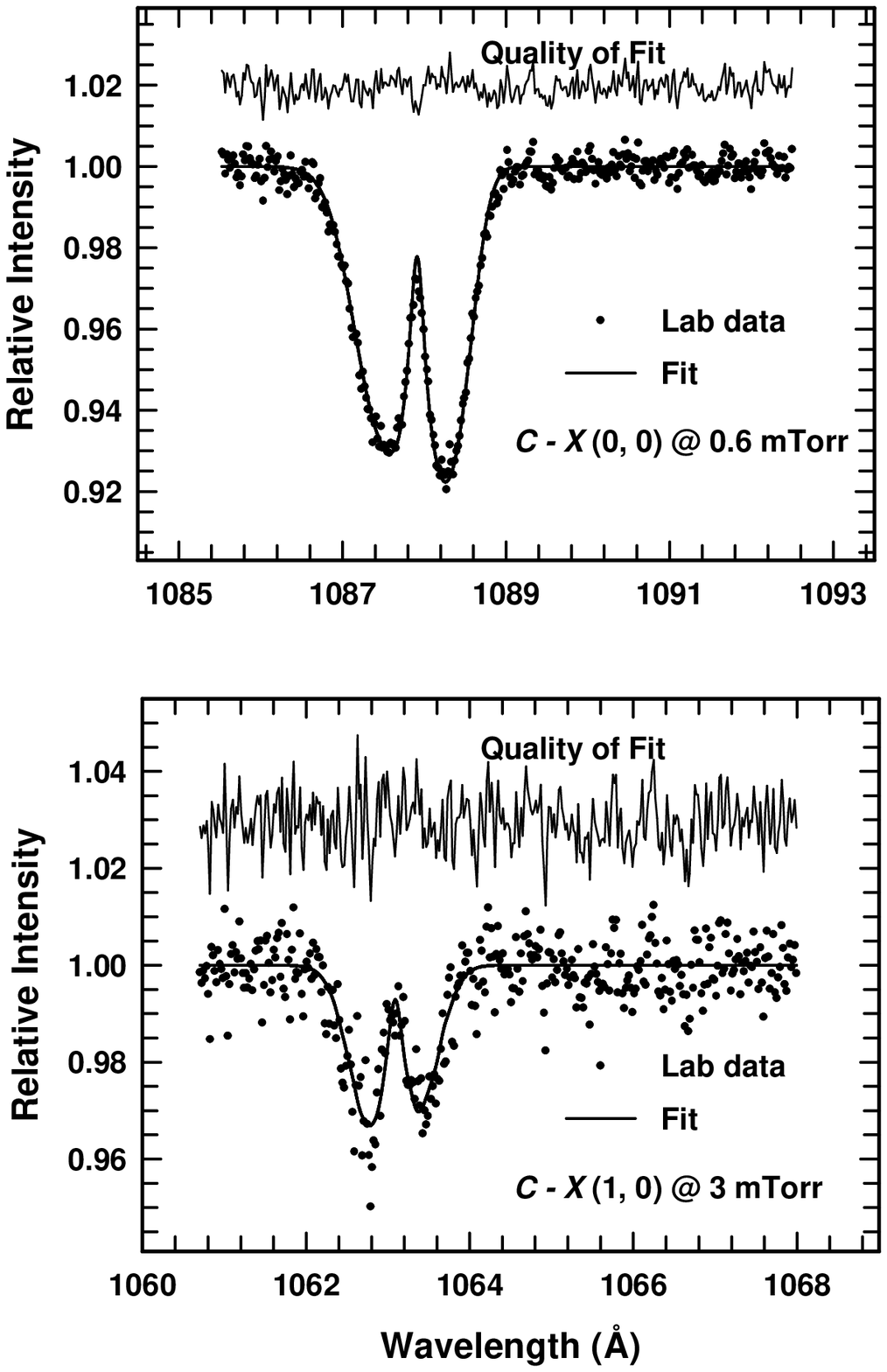}
Fig. 3b
\end{center}

\newpage

\begin{center}
\vbox to5.80in{\rule{0pt}{5.80in}}
\includegraphics{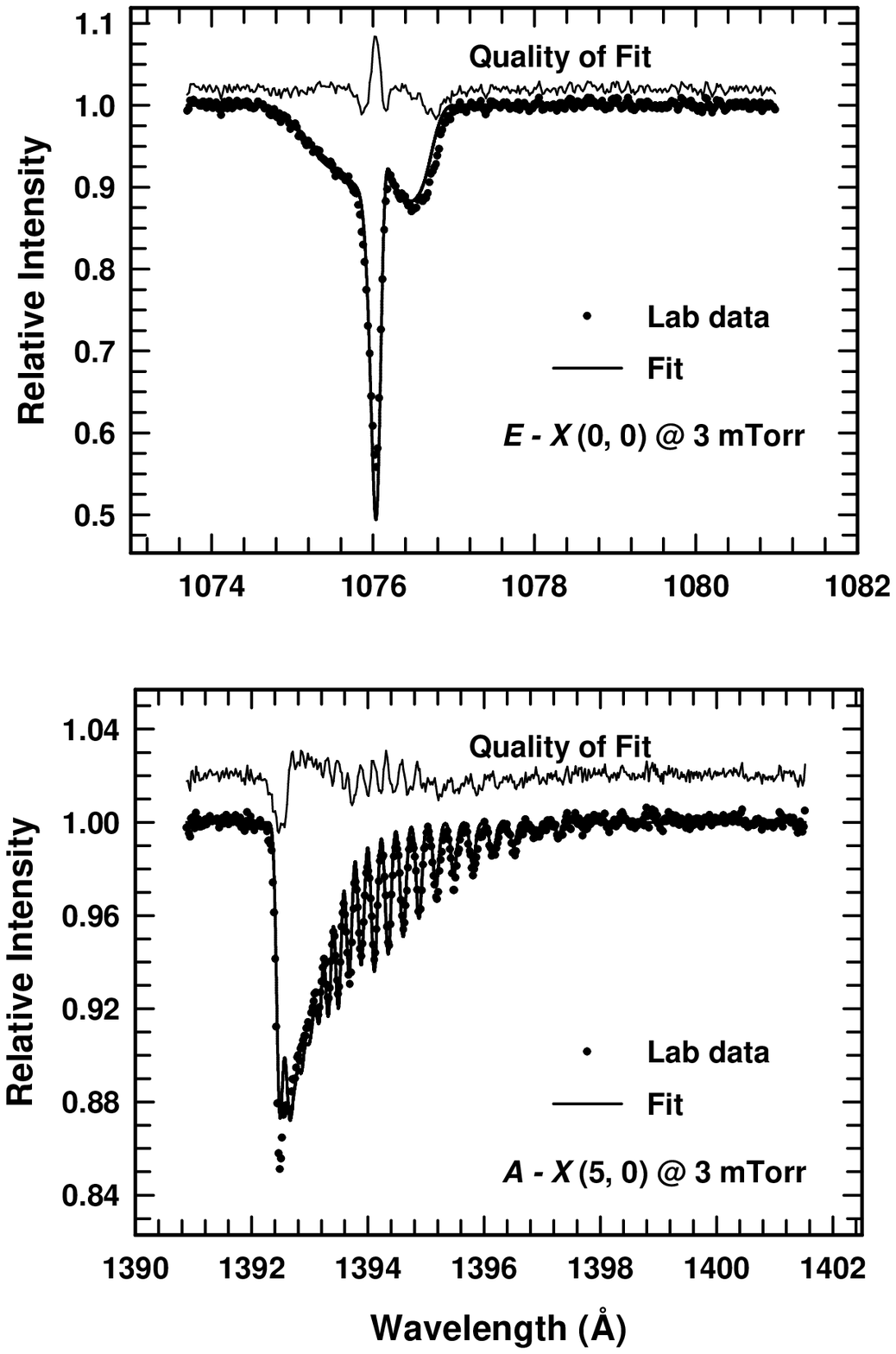}
Fig. 3c
\end{center}

\end{document}